\documentclass[a4paper]{spie}  

\usepackage[]{graphicx}
\usepackage{url}
\usepackage{textgreek}

\title{The LOFT (Large Observatory for X-ray Timing) \\ background simulations} 

\author{R.~Campana\supit{a,b}, M.~Feroci\supit{a,b}, E.~Del~Monte\supit{a,b}, S.~Brandt\supit{c}, C.~Budtz-J\o rgensen\supit{c}\\ N.~Lund\supit{c}, J.~Alvarez\supit{d}, M.~Hernanz\supit{d}, E.~Perinati\supit{e}
\skiplinehalf
\supit{a}INAF/IAPS, Via Fosso del Cavaliere 100, I-00133, Roma, Italy; \\
\supit{b}INFN/Sezione di Roma 2, Via della Ricerca Scientifica 1, I-00133, Roma, Italy;\\
\supit{c}DTU Space, Elektrovej Building 327, DK-2800, Kgs. Lyngby, Denmark;\\
\supit{d}IEEC/CSIC, Campus UAB, E-08193, Bellaterra, Spain;\\
\supit{e}IAAT, Sand 1, D-72076, T\"ubingen, Germany.\\
}

\authorinfo{\emph{On behalf of the LOFT consortium.} \\ Send correspondence to R. Campana. E-mail: \url{riccardo.campana@iaps.inaf.it}}

\begin{document} 
\maketitle 

\begin{abstract}
The Large Observatory For X-ray Timing (LOFT) is an innovative medium-class mission
selected for an assessment phase in the framework of the ESA M3 Cosmic Vision call.
LOFT is intended to answer fundamental questions about the behavior of matter in the very strong gravitational and magnetic fields around compact objects. With an effective area of $\sim$10 m$^2$  LOFT will be able to measure very fast variability in the X-ray fluxes and spectra. A good knowledge of the in-orbit background environment is essential to assess the scientific performance of the mission and to optimize the instrument design. The two main contributions to the background are cosmic diffuse X-rays and high energy cosmic rays; also, albedo emission from the Earth is significant. These contributions to the background for both the Large Area Detector and the Wide Field Monitor are discussed, on the basis of extensive Geant-4 simulations of a simplified instrumental mass model. 
\end{abstract}


\section{Introduction}\label{s:intro} 
An early determination of the background of an X-ray astronomy instrument is an essential feature to drive the detailed design of the detectors and to assess the scientific performance of the experiment. 
The background rate can be estimated by means of a Monte Carlo simulation, in which a somewhat simplified mass model of the spacecraft and of the instruments is specified, and events are generated according to the main features of the radiation environment expected in orbit. For example, in the case of a mission in low-Earth orbit (LEO) the main source of background is due to the interaction of the photon diffuse high-energy radiation and of particles (both cosmic rays and Earth-trapped) with the satellite materials surrounding the instruments.

In this paper we will show Monte Carlo simulations of the instrumental background for the LAD (Large Area Detector) and WFM (Wide Field Monitor) instruments onboard the proposed LOFT (Large Observatory for X-ray Timing) mission\cite{feroci11}.

This paper is structured as follows. In Section \ref{s:loft} we will describe the main instruments onboard the LOFT satellite, the LAD and the WFM. In Section \ref{s:bkg} we will summarize the main contributions to the instrumental background, while in Section \ref{s:mc} the main features of the LAD and WFM Monte Carlo mass models are described. The simulation results are shown in Section \ref{s:results}, and in Section \ref{s:conclusions} we draw our conclusions.

\section{The LOFT mission and its instruments}\label{s:loft} 
The medium-class LOFT mission was selected in 2011 by the European Space Agency for an assessment phase in the framework of the M3 Cosmic Vision program. 
The main goal of this space observatory is to investigate the behavior of the matter in the most extreme physical conditions, that can be found in the gravitational and magnetic fields around neutron stars and black holes. 
LOFT will exploit as a diagnostic the rapid spectral and intensity variability of the X-ray emission from these objects. 

\begin{figure}[htbp]
\centering
\includegraphics[scale=0.5]{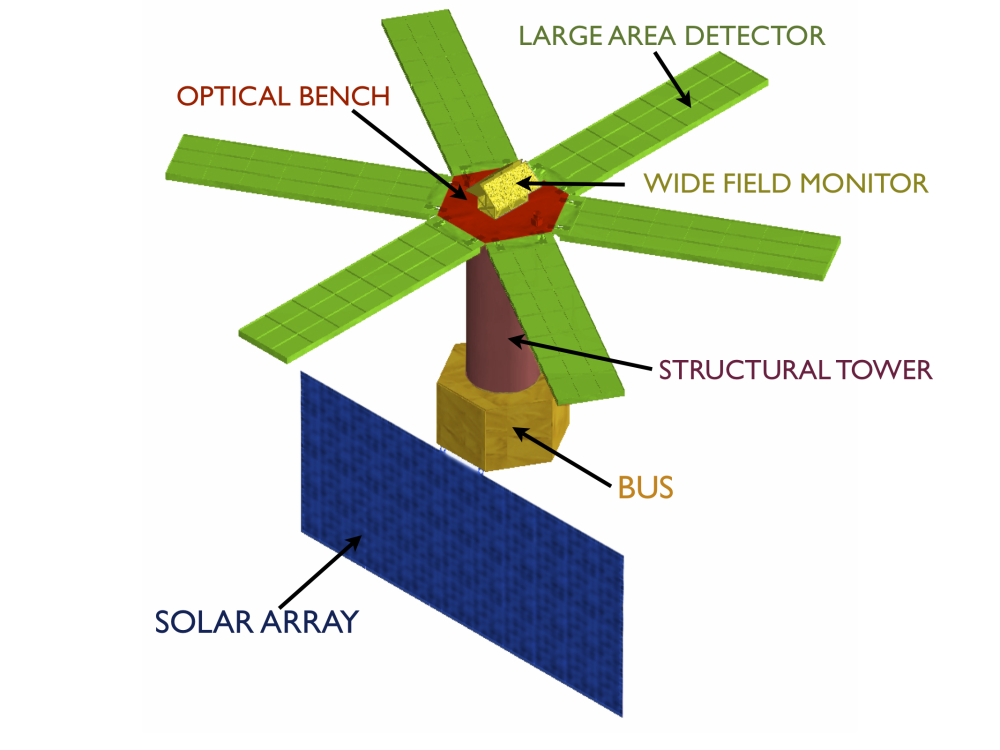}
\caption{The LOFT mission and its instruments.}
\label{fig1}
\end{figure}

The main instrument, the Large Area Detector, has a collecting area 20 times greater than its largest predecessor (PCA onboard RXTE), and is an array of $\sim$2000 individual large-area Silicon Drift Detectors (SDDs), sensitive to the 2--30 keV  emission collected in a $\sim$1$^\circ$ field of view by means of a lead-glass microcapillary collimator plate. 
The instrument design is modular: an array of 4$\times$4 SDDs and their front-end electronics are placed in a \emph{module}, and 21 modules compose a \emph{panel}. The current LAD design envisages 6 panels, that stems from the satellite bus and are deployed once the spacecraft reaches its orbit (Figure \ref{fig1}).
The current baseline design foresees collimators with the same footprint as the SDDs ($\sim$11 cm $\times$ 7 cm), 6 mm thickness, and 100$\mu$m wide square pores separated by 20$\mu$m thick walls (and thus having an \emph{open area ratio}, OAR, of $\sim$70\%).
The main features of the LAD instrument are summarized in Table \ref{t:LADchar}.

\begin{table}[htdp]
\caption{Main characteristics of the LAD instrument.}
\begin{center}
\begin{tabular}{c|c}
Energy range & 2--30 keV (2--80 keV extended range) \\ \hline
Effective area & 10 m$^2$ \\ \hline
Field of view & $\sim$1$^\circ$ \\ \hline
Energy resolution & 200 eV FWHM at 6 keV \\ \hline
Time resolution & 7$\mu$s \\ 
\end{tabular}
\end{center}
\label{t:LADchar}
\end{table}

In order to survey a large fraction of the sky simultaneously, and to trigger follow-up observations with the main instrument, LOFT will host also a coded-mask Wide Field Monitor (WFM). This instrument, sensitive in the same energy range of LAD, will use basically the same SDD detectors, with some design modifications to allow for imaging\cite{campana11, evangelista12, donnarumma12}.
The current WFM design envisages five \emph{units} pointing to different sky directions, each composed by two \emph{cameras}. The ten cameras have the same geometrical dimensions, with a 14.25~$\times$~14.25~cm$^{2}$ detector plane and a 26~$\times$~26~cm$^{2}$ coded mask at a distance of $\sim$20 cm.
Due to the asymmetric spatial resolution of the SDDs\cite{evangelista12}, each camera in a unit will image the same sky direction, but with mutually orthogonal ``fine" resolution directions, thus ensuring both redundancy and a good total angular resolution.
The main features of the WFM instrument are summarized in Table \ref{t:WFMchar}.

\begin{table}[htdp]
\caption{Main characteristics of the WFM instrument.}
\begin{center}
\begin{tabular}{c|c}
Energy range & 2--50 keV (50--80 keV extended)\\ \hline
Geometric area & 2030 cm$^2$ (10 cameras) \\ \hline
Sensitive area & 1820 cm$^2$ (10 cameras) \\ \hline
Field of view (FWZR) & 90$^\circ$ $\times$ 90$^\circ$ (1 camera)  \\ \hline
Energy resolution & $<$300 eV FWHM at 6 keV \\ \hline
Angular resolution &  5$' \times$ 5$'$ (2 cameras) \\ \hline 
On-axis sensitivity (5$\sigma$, 50 ks) & 3.1 mCrab (1 camera) \\ 
\end{tabular}
\end{center}
\label{t:WFMchar}
\end{table}

\section{Sources of background}\label{s:bkg} 
LOFT will be launched by a Soyuz vector in an equatorial low-Earth orbit, at an altitude of $\leq$600 km and with an inclination of $\leq$5$^\circ$. Here, the geomagnetic field effectively screens the primary, high-energy cosmic rays up to energies of a few GeVs.
Moreover, in this low-inclination orbit the South Atlantic Anomaly is only grazed in its external regions. Therefore, the activation of materials in this high-radiation environment is negligible.

The main sources of background considered for the simulations are thus:
\begin{enumerate}
\item \emph{Cosmic diffuse X-ray background}. For the cosmic X-ray and $\gamma$-ray diffuse background we assume the Gruber \cite{gruber99} analytic form, derived from HEAO-1 A4 measurements and valid in the range 3 keV--100 GeV (in photons cm$^{-2}$ s$^{-1}$ sr$^{-1}$):
\begin{equation}
F(E) = \left\{ \begin{array}{ll}
         7.877 \times \left( \frac{E}{1 \mathrm{\, keV}} \right)^{-1.29} e^{-E/41.13} & \mbox{for $E < 60$ keV};\\
        0.0259  \times \left( \frac{E}{60 \mathrm{\, keV}} \right)^{-5.5} +  0.504\times \left( \frac{E}{60 \mathrm{\, keV}} \right)^{-1.58}  +   0.0288  \times \left( \frac{E}{60 \mathrm{\, keV}} \right)^{-1.05} & \mbox{for $E > 60$ keV}.\end{array} \right.
\end{equation}

\item \emph{Earth albedo $\gamma$-rays}. The secondary photon background is due to the cosmic ray (hadronic and leptonic components) interaction with the atmosphere and to the reflection of the CXB. This albedo component has a higher flux, for unit of solid angle, than the CXB for energies above 70 keV. We assumed the albedo spectrum as measured by BAT\cite{ajello08}: this spectrum agrees in the range above 50 keV, after some corrections, with the previous measurements performed by the 1972-076B satellite\cite{imhof76}.
\item \emph{Residual primary cosmic rays}. The geomagnetic cut-off for the considered LOFT orbit lies at rigidities of a few GeVs. An average residual flux of protons, $\alpha$ particles, electrons and positrons has been simulated, using as an input the AMS measurements\cite{alcaraz00a, alcaraz00b} in low-Earth, equatorial regions.
\item \emph{``Secondary" cosmic rays}. In the considered LEO orbit, the impinging particle spectrum consists also of a secondary quasi-trapped component, originating from the primary particles that impacts the Earth atmosphere (sometime they are referred as the ``splash" and ``reentrant" components). We used the proton, positron and electron spectra measured by AMS\cite{alcaraz00a, alcaraz00b} and analytically modelled by Mizuno et al.\cite{mizuno04}.
\item \emph{Earth albedo neutrons}. To account for the flux of neutrons produced by cosmic-ray interactions in the Earth atmosphere, we used the QinetiQ Atmospheric Radiation Model (QARM)\cite{lei04}.
\item \emph{Natural radioactivity}. The lead-glass used in the microcapillary plate collimators contains potassium, in a fraction of approximately 7.2\% by weight. The activity due to the naturally occurring radioactive isotope $^{40}$K is considered, with the emission of $\beta$-rays up to an energy of 1.31 MeV ($\sim$89\% branching ratio) and of 1.46 MeV photons ($\sim$11\% b.r.) due to electronic capture.
\end{enumerate}

In the simulations, photons are isotropically generated along all the possible incoming directions, and then the resulting fluxes are properly normalized for the corresponding exposure time and solid angle.

\section{The LOFT Geant-4 Monte Carlo simulator}\label{s:mc} 
Simulations were performed by using the Geant-4 Monte Carlo toolkit\cite{agostinelli03} (version 4.9.4). Geant allows to describe the geometry and the materials of the instruments and of the satellite bus. Moreover, the code enables to follow the various physical interactions along the path of a primary event through the various components of the geometry, evaluating the secondary particles and the energy deposits generated. 

For the description of the electromagnetic interactions, we used the Low Energy Livermore library\footnote{\url{https://twiki.cern.ch/twiki/bin/view/Geant4/LoweMigratedLivermore}}, that trace the interactions of electrons and photons with matter down to about 250 eV, using interpolated data tables. The hadronic physics, instead, is described using different models in different energy ranges (e.g. Bertini Cascade model, Quark-Gluon String model, etc.). Both elastic and anelastic processes are treated.

\subsection{LAD}
\begin{figure}[htbp]
\centering
\includegraphics[scale=0.3]{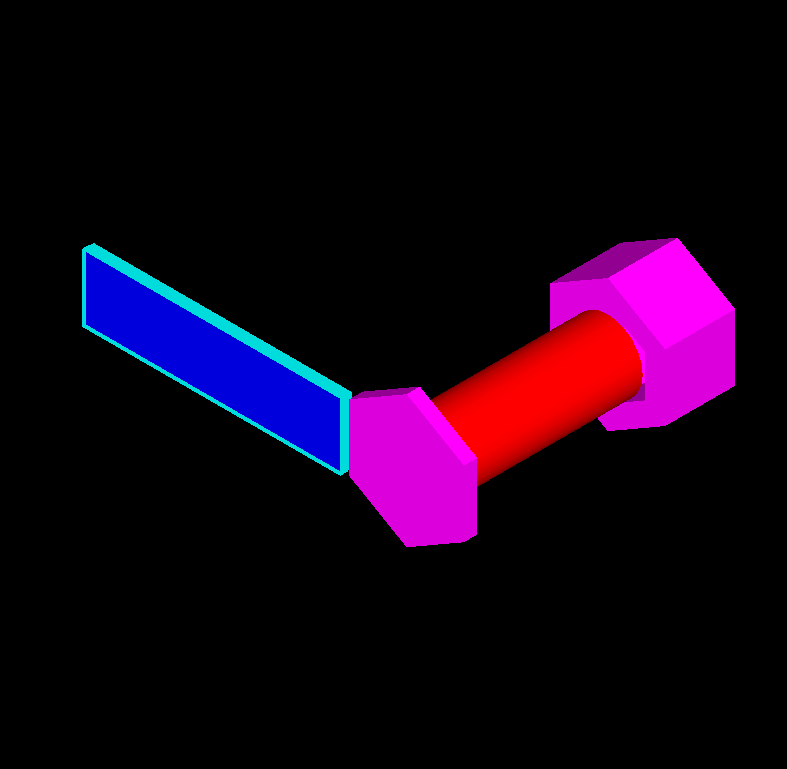}
\caption{The Geant-4 LOFT/LAD mass-model used in the simulations. The panel structure is shown on the left, while the satellite bus is on the right.}
\label{fig2}
\end{figure}

\begin{figure}[htbp]
\centering
\includegraphics[scale=0.3]{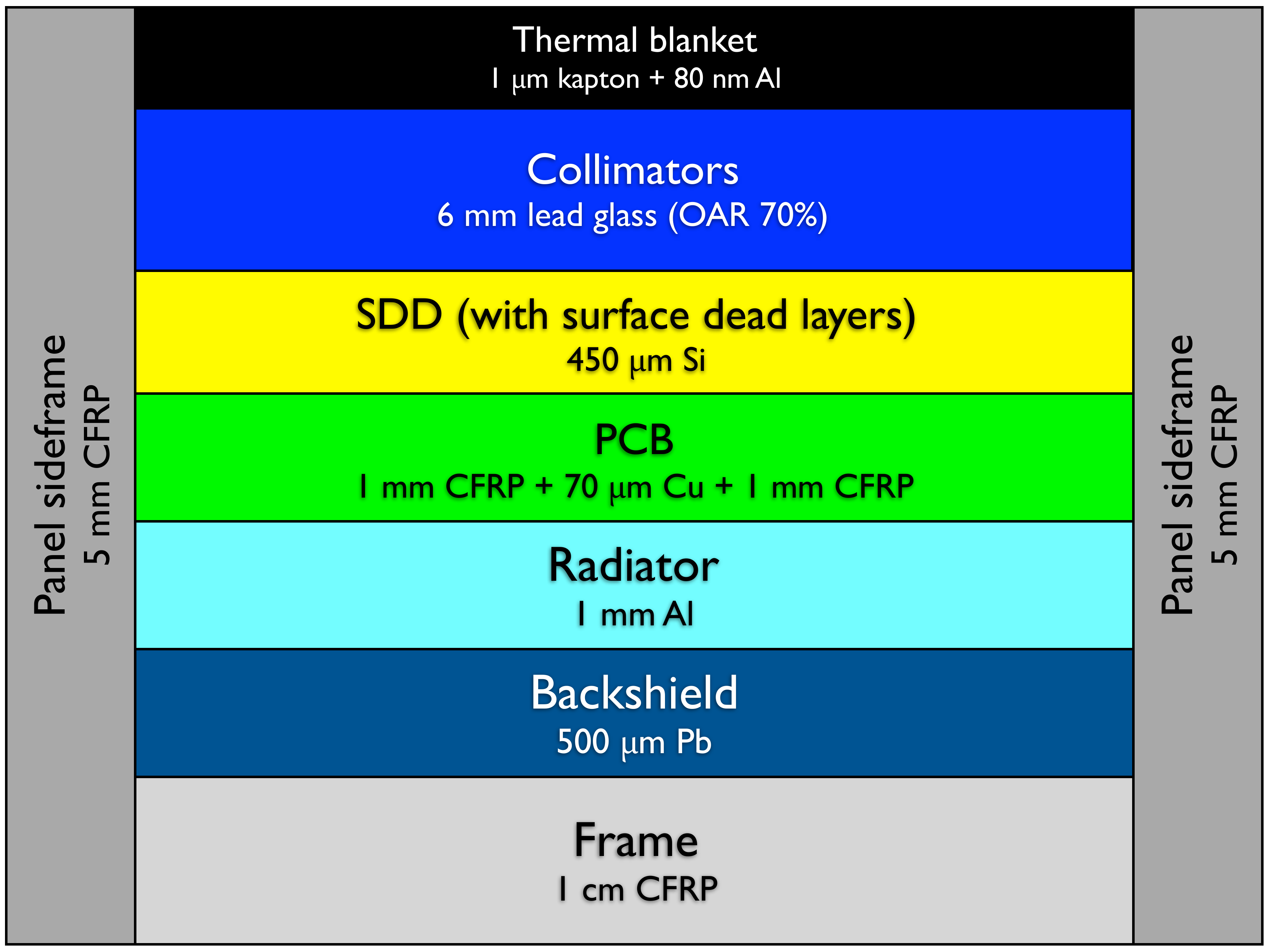}
\caption{Conceptual scheme of the LAD module structure used in the simulations (not to scale).}
\label{ladstructure}
\end{figure}

The mass model geometry used in the simulations for the LAD instrument is shown in Figure \ref{fig2}. A simplified geometry for the satellite bus and structural tower is assumed, using aluminium as the material with an ``effective" density that takes into account the total mass and volume. A single LAD panel is simulated, with the actual dimensions ($\sim$ 350 cm $\times$ 90 cm) but with a simplified design that consists of a stacked-layers geometry (sketched in Figure \ref{ladstructure}). 

The SDD array is represented as a 450$\mu$m thick slab of silicon, subdivided in 1~mm~$\times$~35~mm pixels. On its surface various passive layers are placed (Al cathode implants, SiO$_2$ passivation, undepleted Silicon bulk), the MCP collimator (with an ``effective", reduced density, to take into account the holes without resorting to a more accurate but much more time consuming geometrical description) and the thermal blanket (1$\mu$m Kapton coupled with a 80 nm Al layer).
The FEE board (simulated as a 2 mm thick CFRP slab with an interposed 70$\mu$m Cu conductive layer), a 1 mm thick Al radiator and a 500$\mu$m Pb backshield are placed under the detectors. A carbon-fiber reinforced plastic frame surrounds the sides and the bottom of the panel.

\subsection{WFM}
\begin{figure}[htbp]
\centering
\includegraphics[scale=0.4]{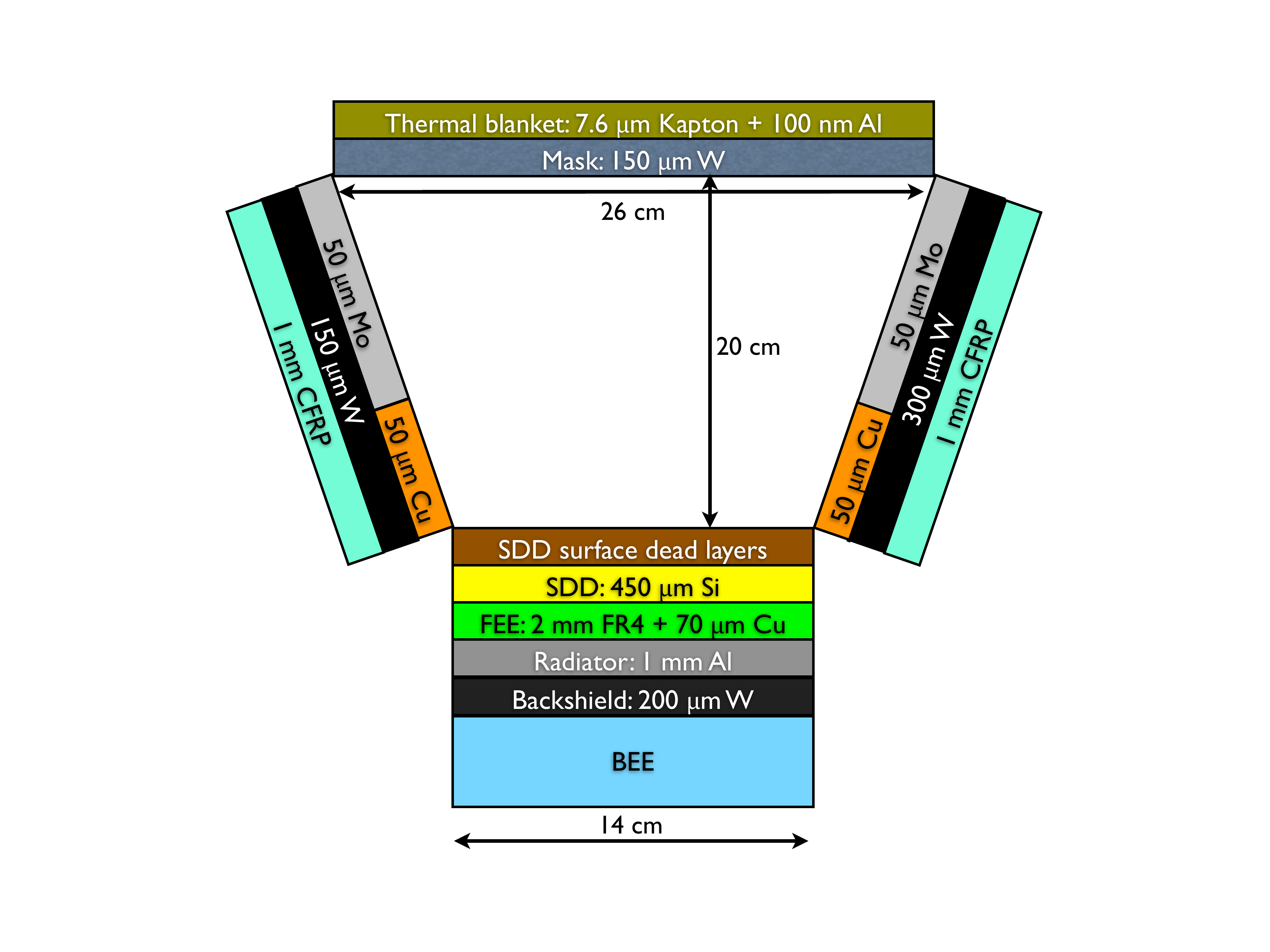}
\caption{Conceptual cross-sectional scheme of the WFM camera geometry and materials used in the simulations (not to scale).}
\label{wfmstructure}
\end{figure}

Presently, only the model of a single WFM camera has been simulated. The geometry is sketched in Figure \ref{wfmstructure}. The tungsten coded mask is 150$\mu$m thick  with 25\% open fraction.
The Silicon Drift Detector, simulated as a 450$\mu$m slab of silicon with 150$\mu$m $\times$ 5 mm pixels, has the usual surface dead layers: Al cathode implants, SiO$_2$ passivation, undepleted Si bulk. The front-end electronics (FEE) board is a sandwich of two 1 mm FR4 layers and a 70$\mu$m Cu conductive layer. Below the detector and the front-end electronics PCB, a ``radiator" layer has been simulated, using Aluminium (1 mm thickness) as the material. In order to generate an adequate number of fluorescence lines for calibration, the lateral collimators (1 mm CFRP + 150$\mu$m tungsten) are partially covered by a 50$\mu$m copper layer on the lower third of the internal surface, and a 50$\mu$m layer of molybdenum in the upper two-thirds of the surface.

\section{Results}\label{s:results} 
For both the LAD and the WFM mass models, the simulations have been performed by generating an isotropic flux of primary events (photons or particles) on a sphere surrounding the experiment, and recording the energy deposits in the detector pixels. The resulting counts have been then properly normalized, taking into account the energy spectra of the simulated background contributions. For the LAD, an anode multiplicity rejection algorithm is implemented to filter out ionization streaks from charged particles, that leave an energy deposit on more than 2 adjacent anodes.

\begin{figure}[htbp]
\centering
\includegraphics[scale=0.6]{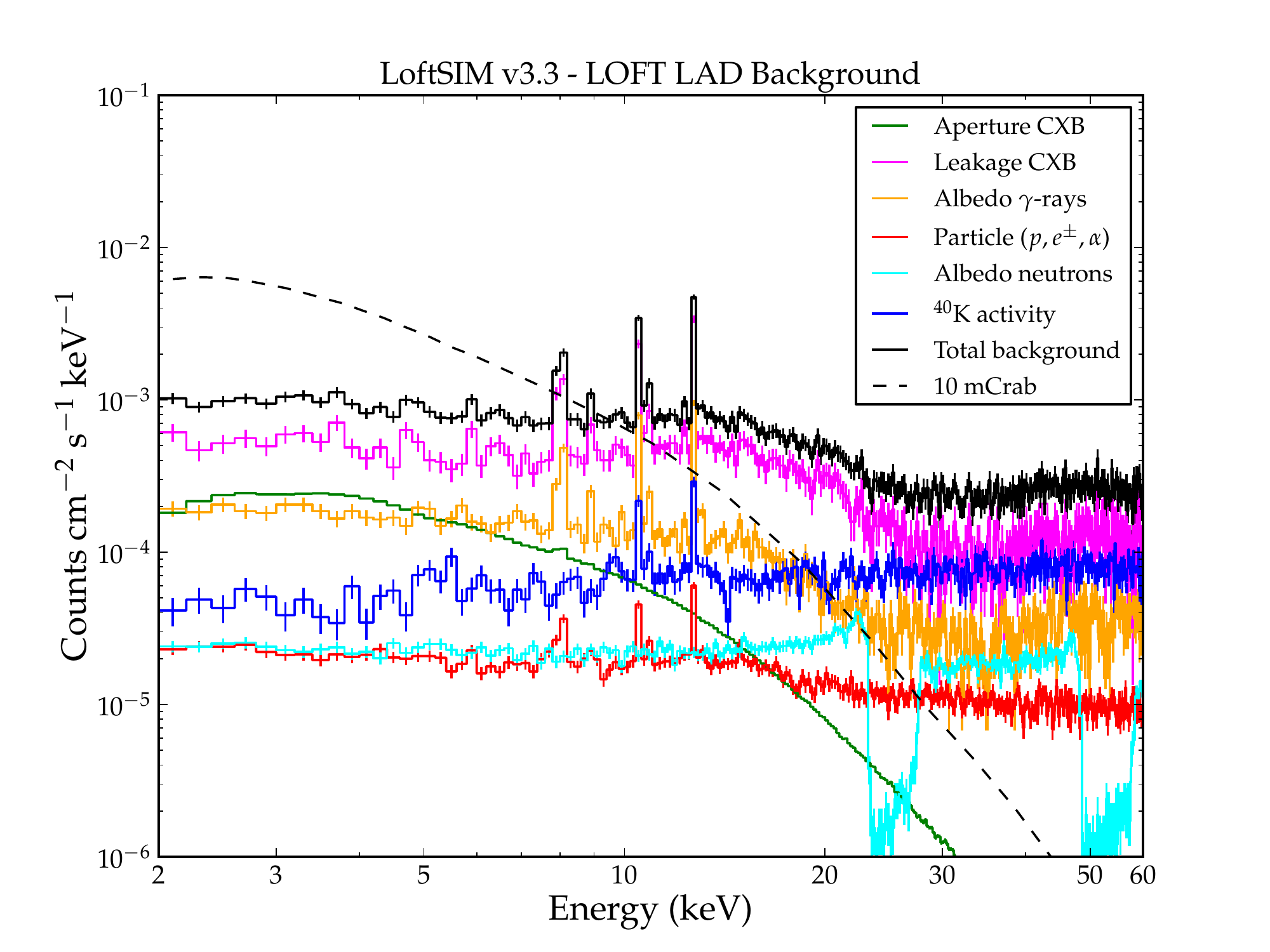}
\caption{The LAD background and its various contributions.}
\label{LADbkg}
\end{figure}

The total resulting LAD background is shown in Figure \ref{LADbkg}. This spectrum has not been smoothed by the instrumental energy resolution (200 eV FWHM at 6 keV for events that are read-out by a single anode).
The main background contribution is due to the CXB photons that leaks from and are scattered in the collimators or in the detector itself, while the diffuse emission collected in the field of view and the particle background are only a minor contribution to the total count rate. Fluorescence emission from the Pb contained in the collimator glass ($L$-shell lines at 10.55 and 12.61 keV) and from the Cu contained in the FEE PCB board ($K$-shell lines at 8.05 and 8.90 keV) is present. Analysis are ongoing to evaluate whether these lines are to be shielded or used as in-flight calibration lines.
In the plot the spectrum of a 10~mCrab point-like source (dashed line) is also shown. 
The total background count rate breakdown is shown in Table \ref{LADbkgtable}.

\begin{table}[htdp]
\caption{The LAD background contributions. The count rates are in the 2--30 keV band.}
\begin{center}
\begin{tabular}{c|c}
Contribution & counts cm$^{-2}$ s$^{-1}$ \\
\hline \hline
Aperture CXB   & 1.8 $\times$ 10$^{-3}$ \\
Leakage CXB   & 1.2 $\times$ 10$^{-2}$ \\
Earth albedo $\gamma$-rays   & 3.5 $\times$ 10$^{-3}$ \\
Earth albedo neutrons   & 5.7 $\times$ 10$^{-4}$ \\
Cosmic-ray particles   & 4.6 $\times$ 10$^{-4}$ \\ \hline
Total background   & 1.8 $\times$ 10$^{-2}$ \\
\end{tabular}
\end{center}
\label{LADbkgtable}
\end{table}%

\begin{figure}[htbp]
\centering
\includegraphics[scale=0.6]{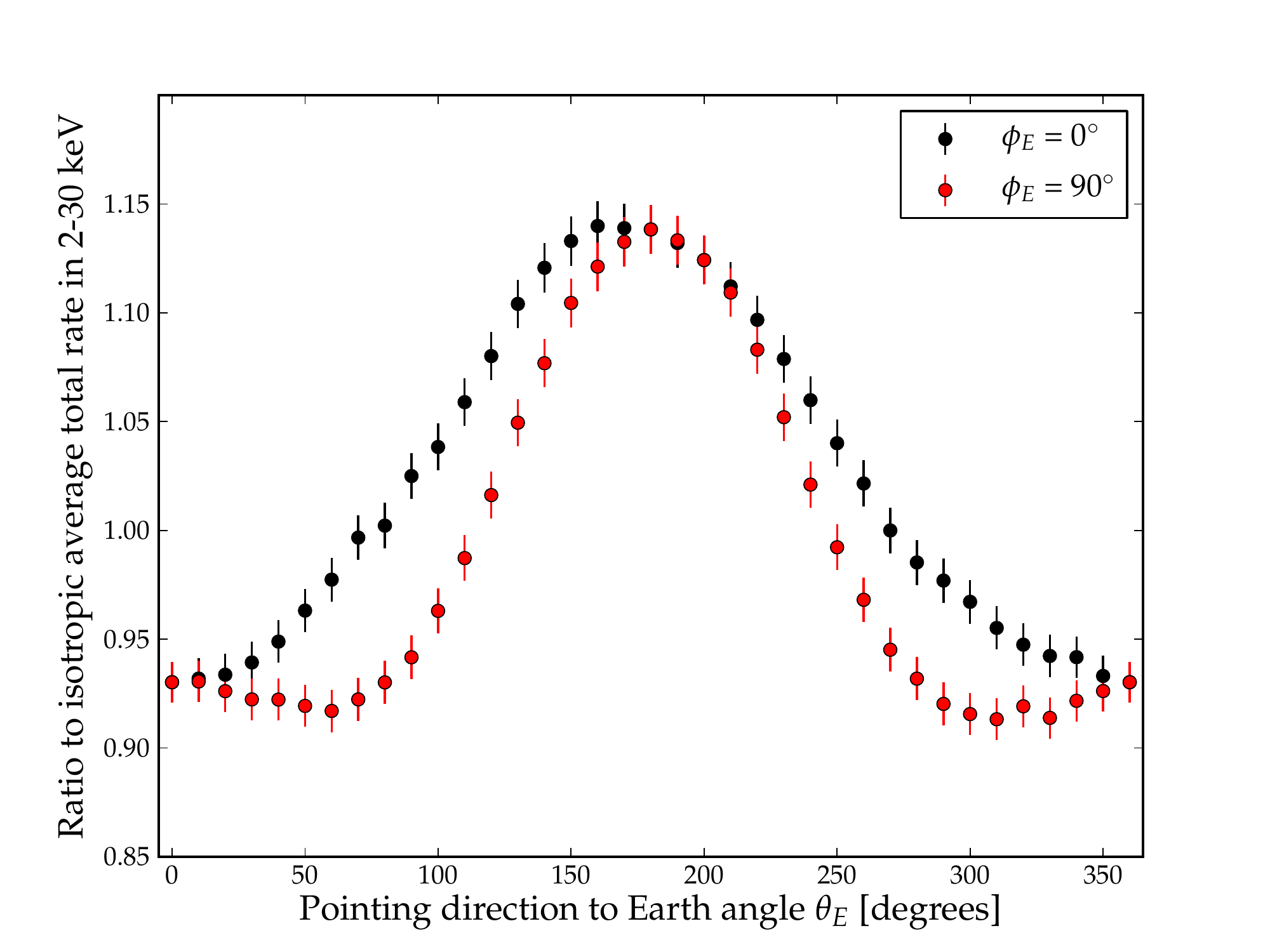}
\caption{Background modulation due to the varying position of the Earth with respect to the pointing direction. $\theta_E$ is the polar angle between the LAD axis and the center of the Earth, $\phi_E$ the azimutal angle.}
\label{bkg_rot_rate_avg}
\end{figure}

The main components of the LAD background shown in Figure \ref{LADbkg} are the leakage through the collimator of high energy photons from CXB emission and the Earth $\gamma$-ray albedo, steady sources but with different intensity and spectra. The orientation of the LAD in this radiation environment determines a small variation of the expected background, due to geometry. We studied this effect, finding that the maximum expected modulation of the background is of about 20\% (as compared to a factor of a few for instruments dominated by particle-induced background).  
Figure \ref{bkg_rot_rate_avg} shows the background as a function of the angle between the LAD pointing direction and the center of the Earth.
$\theta_E = 0^\circ$ corresponds to the Earth center aligned with the field of view, while $\theta_E = 180^\circ$ corresponds to the Earth at the instrument nadir. Moreover, the plot shows the curves also for two values of the azimutal angle: $\phi_E = 0^\circ$ along the short side of the LAD panel, $\phi_E = 90^\circ$ along the long side of the LAD panel (the broadening is due to the scattering on the satellite bus structures).
The modulation is fully geometrical and it can be predicted and modeled. However, to improve further on the modeling, we plan to use a set of ``blocked'' detectors aimed at monitoring the background at any time. Simulations show a control on the background better than 0.3\%.

\medskip

\begin{figure}[htbp]
\centering
\includegraphics[scale=0.6]{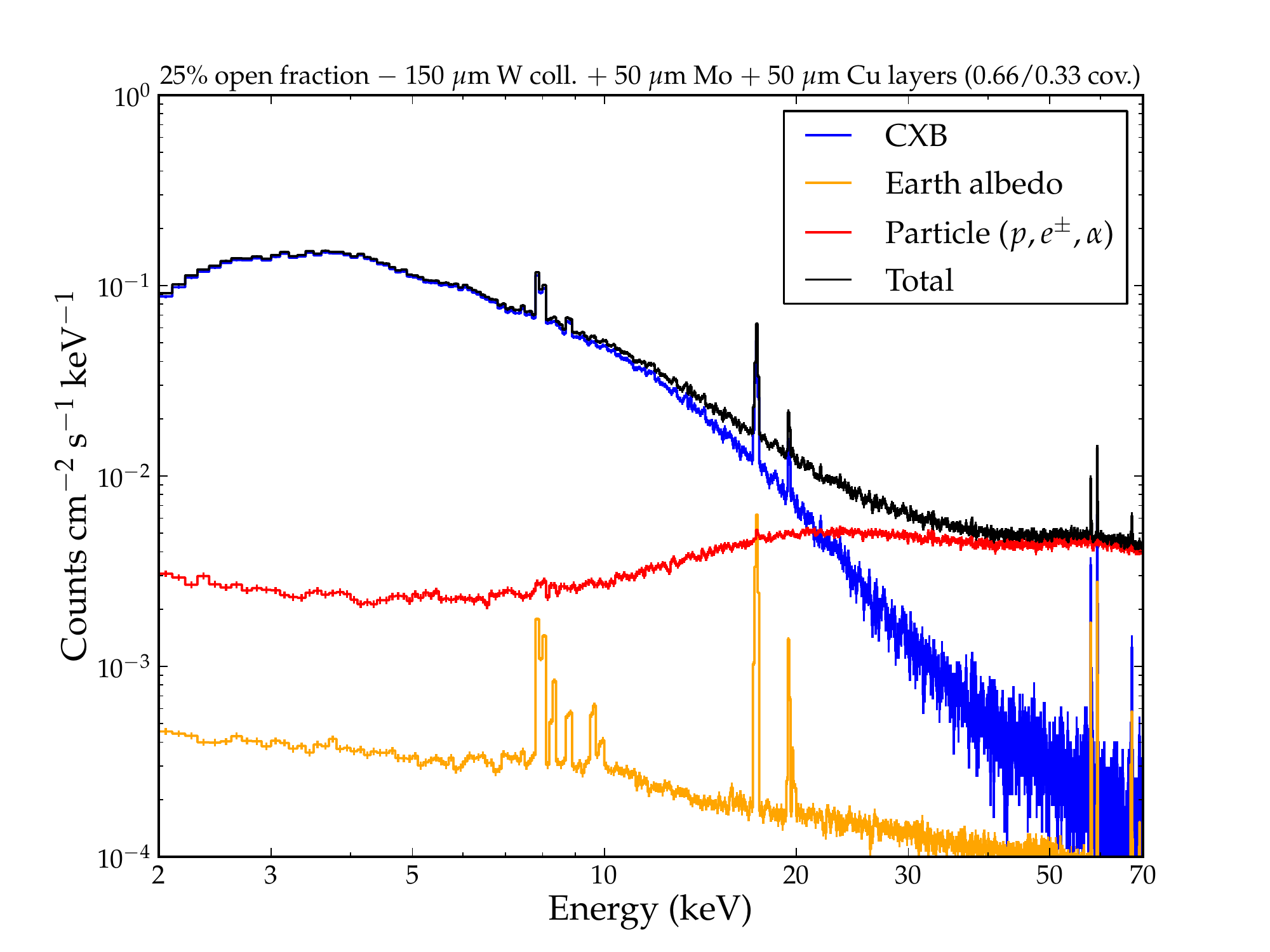}
\caption{The WFM background.}
\label{WFMbkg}
\end{figure}

The total resulting WFM background is shown in Figure \ref{WFMbkg}.
Also in this case the spectrum is not smoothed with the energy resolution. 
Here, due to the large field of view, the main contributions to the background is given by the cosmic diffuse X-ray emission, 
as expected, while the particle background is a minor contribution. In the spectrum the prominent $K$-shell fluorescence lines from Cu (8.05 and 8.90 keV), Mo (17.5 and 19.6 keV), and W (59.3 and 67.2 keV) are noteworthy, and will allow to monitor the instrumental calibration on orbital timescales.

\begin{table}[htdp]
\caption{The WFM background contributions. The count rates are in the 2--50 keV band.}
\begin{center}
\begin{tabular}{c|c}
Contribution & counts cm$^{-2}$ s$^{-1}$ \\
\hline \hline
Cosmic X-ray diffuse background   & 1.0 $\times$ 10$^{0}$ \\
Earth albedo $\gamma$-rays   & 1.1 $\times$ 10$^{-2}$ \\
Cosmic-ray particles   & 2.0 $\times$ 10$^{-1}$ \\ \hline
Total background   & 1.2 $\times$ 10$^{0}$ \\
\end{tabular}
\end{center}
\label{WFMbkgtable}
\end{table}%

\section{Conclusions}\label{s:conclusions} 
A simplified mass model of the LAD and WFM instruments onboard LOFT has been developed in order to assess the expected level of instrumental background. 

For the LAD experiment the main contribution to the overall background is due to the diffuse hard X-ray photons leaking from the collimator. This property allows to reach a high degree of control on the background modeling. Particle-induced background is found to be a minor contribution, mainly because of the favourable low-Earth orbit.

A WFM camera, instead, thanks to its large field of view will directly collect the diffuse X-ray background. The interaction with the collimator materials will also produce fluorescence lines that could be used to monitor the instrument calibration. Particle-induced background is a minor contribution at low energies, and becomes dominant only above 20--30 keV.

A detailed discussion of the Monte Carlo simulations and of the resulting background spectra will be the subject of a forthcoming paper (Campana et al., in preparation).

\bibliography{bibliography} 
\bibliographystyle{spiebib}  

\end{document}